\documentclass[epj]{svjour}

\usepackage{graphics}
\usepackage{graphicx}
\usepackage{dcolumn}
\usepackage{bm}

\usepackage{epsfig}
\usepackage{amsmath}
\usepackage{amstext}
\usepackage{amssymb}

\newcommand{\beq}{\begin{eqnarray}}
\newcommand{\eeq}{\end{eqnarray}}

\begin{document}

\title{Limit distributions of scale-invariant probabilistic models of correlated random variables with the $q$-Gaussian as an explicit example}

\author{Rudolf Hanel\inst{1} \and Stefan Thurner\inst{1,3} \and Constantino Tsallis\inst{2,3}}

\institute{                    
  \inst{1} Complex Systems Research Group, HNO, Medical University of Vienna, W\"ahringer G\"urtel 18-20, A-1090, Austria\\
  \inst{2} Centro Brasileiro de Pesquisas F\'isicas, Rua Xavier Sigaud 150, 22290-180 Rio de Janeiro-RJ, Brazil\\
  \inst{3} Santa Fe Institute, 1399 Hyde Park Road, Santa Fe, NM 87501, USA
}

\abstract{Extremization of the Boltzmann-Gibbs (BG) entropy 
$S_{BG}=-k\int dx\,p(x) \ln p(x)$  under appropriate norm and width constraints yields the Gaussian distribution 
$p_G(x) \propto e^{-\beta x^2}$. Also, the basic solutions of the standard Fokker-Planck (FP) 
equation (related to the Langevin equation with {\it additive noise}), 
as well as the Central Limit Theorem attractors, are Gaussians. The simplest 
stochastic model with such features is $N \to \infty$ {\it independent} binary random variables, 
as first proved by de Moivre and Laplace.  
What happens for strongly correlated random variables?
Such correlations are often present in physical situations as e.g. systems with long range 
interactions or memory. Frequently $q$-Gaussians, 
$p_q(x) \propto [1-(1-q)\beta x^2]^{1/(1-q)}
\,\,  [p_1(x)=p_G(x)]$ become observed.  
This is typically so if the Langevin equation includes {\it multiplicative noise}, 
or the FP equation to be {\it nonlinear}. 
Scale-invariance,
i.e. exchangeable binary stochastic processes, allow a systematical  
analysis of the relation between correlations and non-Gaussian distributions.
In particular, a generalized stochastic model yielding $q$-Gaussians for all ($q \ne 1$) was missing. 
This is achieved here by using the Laplace-de Finetti representation theorem, which embodies strict 
scale-invariance of interchangeable random variables.  
We demonstrate that strict scale invariance together with $q$-Gaussianity 
mandates the associated {\it extensive} entropy to be BG. 
} 

\maketitle
 
One of the cornerstones of statistical mechanics is the functional connection 
of the thermodynamic entropy with the set of probabilities $\{p_i\}$ of microscopic 
configurations. For the celebrated 
Boltzmann-Gibbs (BG) theory, this  functional is given by
\begin{equation}
S_{BG}=-k \sum_{i=1}^N p_i \ln p_i   \quad, 
\label{BGentropy}
\end{equation}
where $N$ is the total number of microscopic states which are compatible 
with the information available about the system. This powerful connection 
is in principle applicable to a vast class of relevant systems, including (classical) 
dynamical ones whose maximal Lyapunov exponent is positive warranting 
strong chaos, hence mixing in phase space, hence ergodicity 
(Boltzmann, in some sense, embodied all these features in his {\it molecular chaos hypothesis}).  
Within this theory exponential distributions, 
i.e. exponential statistical factors, emerge naturally.

In particular, the Gaussian distribution
$ p_G \propto e^{-\beta x^2} $ with $\beta >0$ 
is found to be the
(i)
velocity (Maxwell) distribution 
of any classical many-body Hamiltonian system in thermal equilibrium with a thermostat
i.e., if the interactions between its elements are sufficiently short-ranged, or inexistent.
Furthermore, 
this important probabilistic form (ii) maximizes the (continuous version of the) 
entropy $S_{BG}=-\int dx \, p(x) \ln [p(x)]$ under the basic constraints of normalizability 
and finite width; (ii) constitutes the exact solution
of the simplest form of the (linear) Fokker-Planck equation 
(i.e. corresponding Langevin equations with purely additive noise); 
(iii) is the $N \to\infty$ 
attractor of the appropriately centered and scaled sum of $N$ {\it independent} (or 
weakly correlated in an appropriate sense) discrete or continuous random variables whose 
second moment is finite (Central Limit Theorem).
%
%
The simplest probabilistic 
model which realizes these paradigmatic features is a set of $N$ independent equal 
binary random variables (each of them taking, say the values 0 and 1, with probability 1/2). 
The probability of having, for fixed $N$, $n$ $1$'s is given by 
$\frac{N!}{n! \,(N-n)!}\, 2^{-N}$. Its limiting distribution is, after centering and scaling, 
a Gaussian (as first proved by de Moivre and Laplace), 
and its (extensive) entropy is the one of $BG$, since $S_{BG}(N)=N k\ln 2$.   

What happens with the above properties
when these binary random variables are not independent and the correlations 
between those variables (homogeneously spread over the system) are strong enough?
There is in principle no reason for 
expecting the relevant limiting distribution to be a Gaussian, and the entropy 
which is extensive (i.e.,  $S(N) \propto N$ for $N \gg 1$) to be $S_{BG}$. 
The purpose of this paper is to answer such questions for a  mathematically intriguing
class of correlated processes -- which can be interpreted as prototypical mean field models -- 
and therefore are relevant in natural, artificial and even social systems.
Let us  discuss the frequently occurring $q$-Gaussians, a natural generalization of the Gaussians, 
defined as
\begin{equation}
p_q(x) \propto [1-(1-q)\beta x^2]^{1/(1-q)} \quad [p_1(x)=p_G(x)] \quad ,
\end{equation}
where
$x \in R $ for $q<3$ (for $q \ge 3$, normalizability is lost), and $x^2 \le 1/[(1-q) \beta]$ for $q<1$. 
%
%
(i) $q$-Gaussians appear, e.g., as the exact solutions of paradigmatic non-Markovian 
Langevin processes \cite{AnteneodoTsallis2003,FuentesCaceres2008} which lead to 
inhomogeneous linear \cite{Borland1998}, or homogeneous nonlinear 
\cite{PlastinoPlastino1995,TsallisBukman1996}
Fokker-Planck equations. 
(ii) $q$-CLT attractors are $q$-Gaussians \cite{UmarovTsallisSteinberg2008}.
(iii) The extremization of $S_q$ with norm and finite width constraints 
yields 
$q$-Gaussians. 
$S_q$ \cite{Tsallis1988} is a (generically {\it nonadditive} \cite{Penrose1970}) 
generalization of BG entropy,  
namely  
\begin{equation}
  S_q = k \frac{1-\int dx\, [p(x)]^q }{ q-1} \;\;\;\; (q \in{R}; \, S_1=S_{BG}) .
\label{qentropy}
\end{equation}
Usually the extensive entropy of systems is given by the Boltzmann Gibbs entropy $S_{BG}$ which 
is identical with $S_q$ for $q=1$. 
%
It is noteworthy at this point that
there
exist systems with 
distribution functions (which are in general not q-Gaussian) depending on the system size 
$N$ in such a way that 
$S_{BG}$ is not the extensive entropy any more while 
$S_{q}$ becomes the {\it extensive} entropy for specific values of $q<1$ denoted by $q_{\rm ent}$, 
i.e., $S_{q_{\rm ent}}(N) \propto N \;\;(N \gg 1)$, \cite{TsallisGellMannSato2005,CarusoTsallis2008}.
As is well known, for all standard short-range-interacting many-body Hamiltonian systems, 
we have $q_{\rm ent}=1$, i.e. 
$S_{BG}$, which is 
identical to $S_{1}$, is extensive. 
(iv) Numerical indications \cite{PluchinoRapisardaTsallis2007} for the distributions of velocities in quasistationary states of long-range Hamiltonians \cite{AnteneodoTsallis1998} suggest $q$-Gaussians. Different interpretations of the situation are given in
\cite{yamaguchi,bouchet}. 
(v) Further, experimental and observational evidence for $q$-Gaussians exists for the 
motion of biological cells \cite{UpadhyayaRieuGlazierSawada2001,cellmove}, 
defect turbulence \cite{DanielsBeckBodenschatz2004}, solar wind 
\cite{BurlagaVinas2005,BurlagaVinasAcuna2006}, cold atoms in dissipative 
optical lattices  \cite{DouglasBergaminiRenzoni2006}, dusty plasma \cite{LiuGoree2008}, 
among others. Numerical indications are also available at the edge of chaos of unimodal 
maps \cite{TirnakliBeckTsallis2007}. 
In \cite{RodriguezSchwammleTsallis2008} a specific model for correlated binary random variable was shown to converge
to $q$-Gaussian distributions in the thermodynamic limit for $q<1$. Yet, no mechanism was given
of how such models can be generated from general principles. This will be done in this paper.
In particular, it will be shown how the model given in \cite{RodriguezSchwammleTsallis2008}
can be generated and how by the same means $q$-Gaussian models can be generated for $1<q<3$.  

In the following we consider binary exchangeable sto\-cha\-stic processes, with correlated elements say from $x\in\{0,1\}$. 
Exchangeable means that the  $N$-point probabilities \\Ê
$p_N(x_1,x_2,\dots,x_N)$ of the process 
are totally symmetric in its arguments for all $N$ and that $p_N$ can always be obtained by
marginalization of $p_{N+1}$. 
In particular, the probability of a {\em specific} sequence $(x_1,x_2,\dots,x_N)$, a microstate,
does not depend on the order of binary events, but only on the number 
$n$ of events in the state $x_i=0$ and $N-n$ events in the state $x_j=1$.
Following the notation in \cite{RodriguezSchwammleTsallis2008} we denote 
this probability as $r^N_n$. 
There are $\binom{N}{n}$ such micro-states (where the order of the $x_i$ is exchanged), 
and  the probability of finding {\em any} situation with $n$ events in one state is 
given by  
\begin{equation} 
p^N_n=\binom{N}{n}r^N_n \quad, \quad \sum_{n=0}^N p^N_n=1 \quad.
\label{prob_p}
\end{equation}
Total symmetry of the micro-states, $\forall N$, implies   
$r^{N-1}_n=r^{N}_n+r^{N}_{n+1}$, sometimes referred to as the 
{\it Leibniz triangle rule} 
\cite{TsallisGellMannSato2005}, or {\it scale-invariance} of the distribution.
Scale-invariance means here that at every 'scale' $N$ the relation
$p_{N}(x_1,x_2,\dots,x_{N})= \sum_{x_{N+1}} \, p_{N+1}(x_1,x_2,\dots,x_{N+1})$ 
holds (where the sum runs over the possible discrete values of the random variable), 
i.e. the distribution at a lower scale can always be obtained by marginalization of higher scales.
What does this mean for physical systems?
If we look at the binary stochastic process in terms of coin tosses, this means that
the transition probabilities that bias the $N$'th coin toss depend only on the number
of tails and heads thrown after $N-1$ tosses and not on the particular history of coin tosses.
If the coin tosses steer a random walk this means that the next step of the random walk is biased by 
the elapsed time (number of coin tosses $N$) and the distance the random walk has covered to its origin ($N-2n$) 
but not on the particular path the random walk has taken. 
% ...
In all Hamiltonian systems where the interactions are sufficiently long-range and strong enough for mean field solutions to 
be exact or close to exact totally symmetric correlations become physically relevant. Clearly, when mean field solutions become 
exact correlations between the state variables become totally symmetric. 
% ...
For binary state variables, i.e. Ising like spin systems,
this implies that the conditional probability of one arbitrary spin 
being in one particular state conditioned on the state of all other spins only depends on the number of spins in one or the other 
state and not on the particular spin configuration of the system. Correlations introduced by such interactions can be understood
in terms of exchangeable processes. 

For dealing with such binary correlated systems
the following representation was suggested by Laplace in 1774, and later rediscovered by de Finetti \cite{definetti,jaynes};
\begin{equation}
r^N_n=\int_0^1 dy\,y^n(1-y)^{N-n}g(y) \;\;\;\;\; \Bigl[\int_0^1 dy\,g(y)=1 \Bigr] \,.
 \label{definetti}
\end{equation}
This representation ensures Leibniz triangle rule  and normalization,  $\sum_{n=0}^N \binom{N}{n}r^{N}_n=1$. 
Note that the  non-negative $g$ introduces the correlations in the stochastic process.
In the case of independent variables, we have  $g(y)= \delta(y-p)$, i.e.
$r^N_n=p^n(1-p)^{N-n}$ is the usual binomial factor for independent trials.
The fact that {\em any} stochastic process which is exchangeable to all scales $N$ 
can be represented in this way is called  the Laplace-de Finetti theorem \cite{definetti,hewittsavage,jaynes} 
(if processes are exchangeable only up to a scales $N_{max}$ positivity of $g$ is not ensured and negative
correlations can occur \cite{jaynes}).

In order to pass from probabilities to distributions and from sums to integrals
in a Riemann-Stieltjes sense in the thermodynamic limit $N\to\infty$ we
use the well known properties of the Beta function, 
$B(a,b)\equiv\int_0^1 dx\, x^{a-1}(1-x)^{b-1}$, to re-express the binomial factor 
in Eq. (\ref{prob_p}). 
Defining $\rho^N_n \equiv (N+1)p^N_n$ turns Eq. (\ref{prob_p}) into
\begin{equation}
\rho^{N}_n=\frac{\int_0^1 dx\,x^n(1-x)^{N-n}g(x)}{\int_0^1 dx\,x^n(1-x)^{N-n}}
\;,\quad 1=\sum_{n=0}^N\frac{1}{N+1}\rho^N_n\quad.
\end{equation}
As a consequence of this construction the natural domain
of the limit distribution $\rho(y)$ is the interval $y\in[0,1]$.
Using $y^N_n=(1/2+n)/(N+1)$, for $n=0,1,\dots,N$, as discretization of $[0,1]$
in the limit one gets $1/(N+1)\to dy$ and $y^N_n\to y$.
Denote the natural number closest to the value $(N+1)y-1/2$ by $[(N+1)y]$ then the limit distribution gets $\rho(y)=\lim_{N\to\infty}\rho^N_{[(N+1)y]}$.
These arguments imediately lead to
\begin{equation}
\rho(y)= \lim_{N\to\infty}\frac{\int_0^1 dx\,\left[x^y(1-x)^{1-y}\right]^Ng(x)}{\int_0^1 dx\,\left[x^y(1-x)^{1-y}\right]^N}\quad.
\label{eqa}
\end{equation}
Notice that the maximum  of $x^y(1-x)^{1-y}$ with respect to $x$ is obtained at $x=y$ and gets 
amplified above all other values of the function by the power $N$. 
For $N\to\infty$ we therefore get 
a delta sequence and
\begin{equation}
\delta(x-y)= \lim_{N\to\infty}\frac{\left[x^y(1-x)^{1-y}\right]^N}{\int_0^1 dx\,\left[x^y(1-x)^{1-y}\right]^N}\quad. 
\label{eqb}
\end{equation}
Consequently, 
\begin{equation}
\rho(y)= \int_0^1 dx\,\delta(x-y)g(x)=g(y)\quad,
\label{eqc}
\end{equation}
i.e. $\lim_{N\to\infty}\sum_n|\rho^N_n - g(y^N_n)|=0$. 
The limit distribution $\rho$ and the function $g$ generating the 
Leibniz-triangle 
$r^N_n$ are in fact identical!
Therefore, once a desired limiting distribution on $[0,1]$ is given, 
one can simply write down the sequence of $r^N_n$ which is generating it!
Though it is known that $g$ may become negative \cite{jaynes}, 
if scale-invariance is broken and the model is exchangeable only up to a maximal scale $N_{max}$,
this interpretation of $g$ makes it clear why non-negativity of $g$ is required in the limit $N\to\infty$.
The limit distribution $\rho(y)$, i.e. $g(y)$, of binary exchangeable processes
is defined on $y\in[0,1]$, where $y$ is basically the ratio of events $n/N$ in the 
limit of large $N$. 
However, prototypical processes, e.g. spin-systems or random walks, depend
on binary processes (spins up/down or number of left/right steps). 
Yet, their associated observable variable is not the ratio $n/N$ of binary events but 
some other descriptive variable $z$ (the magnetization or the position of a random walk).
Since the domain of the descriptive 
variable $z$ will in general not be $[0,1]$ (e.g. typically $[-1,1]$ for spin systems and $[-\infty,\infty]$ for random walks) it has to be clear from the beginning that the form of the limit distribution of the descriptive variable $z$ will not only depend on $g$
but also on the relation between $z$ and the binary process, i.e. on $y=n/N$, which may or may not depend on the scale $N$. 
E.g., if equidistant spacing on intervals $[-R_N, R_N]\subset[-\infty,\infty]$ as $R_N\to\infty$ is considered (as is the case of CLT) this relation will in general be a variable transformation depending on the scale $N$. 
For simplicity  we only consider the case of 
$N$ independent transformations of variables between $y$ and $z$.  
We assume a symmetric distribution $G(z)$ of the descriptive variable $z$
that can be obtained by a transformation of variables defined by $dz G(z)= dy \rho(y)$.
We will call a strictly monotonous increasing antisymmetric functions $f$,
such that the effective stochastic variable $z$ will take the values 
$z^N_n=f(2y^N_n-1)$, a symmetric representation of the binary process. 
To be clear, $z=f(2y-1)$ is precisely the change of variables relating the generating function 
$g$ of the binary process with the limit distribution $G$ of the observable process $z$.
In particular, $G$ and $g$ are one-to-one related by $2f'(2y-1)G(f(2y-1))=g(y)$.
Each pair $(f,g)$ exactly defines the distribution $G$ of the binary process.
Moreover, fixing $G$ and the representation $f$ uniquely determines $g$.
Inversely, for a given observable distribution $G$, any pair $(f,g)$ that represents $G$ can serve as a stochastic model for $G$. 
A physical view on the meaning of $f$ and $g$ can be given by Galton's board, where $f$ corresponds to the positions of needles on the board while $g$ fixes the probabilities 
of balls being reflected to the left or right at the condition of hitting
some needle.

We now proceed to derive a stochastic model with $q$-Gaussian limit distributions. 
For the case of $q<1$, the $q$-Gaussian $G_q(z)=\frac{1}{Z_q} [1-(1-q)z^2]^{1/1-q}$ is 
defined on a compact support, $|z|\leq\frac{1}{\sqrt{1-q}}$.   
To identify it with the limiting distribution we have to map $z$ to the unit interval  by 
$z=f(2y-1)=\frac{1}{\sqrt{1-q} } (2y-1)$.
Since the support of the $q$-Gaussian $G_q(z)$ for $q<1$ is compact 
an affine transformation is the natural
choice for the change of variables.  
Under an affine variable transformation $G_q(z)\to\rho(y)= \frac{1}{Z_q} 4 y^{ \frac{1}{1-q} }(1-y)^{\frac{1}{1-q} }$, where 
$Z_q = 4 B\left(\frac{1}{1-q}+1,\frac{1}{1-q}+1\right)$.
Consequently, by introducing the notation $\nu\equiv \frac{1}{1-q}+1$, we get  
$g(y)= \frac{ [y(1-y)]^{\frac{1}{1-q} }}{ B ( \nu , \nu ) }$ 
and, finally,  by using Eq. (\ref{definetti}),
\begin{equation}
   r^N_n(q)=\frac{B\left(\nu+n,\nu+N-n\right)}{B (\nu,\nu )} \quad  .
 \label{result1}
\end{equation}
To retrieve the original $q$-Gaussian $G_q(z)$ one has to perform the inverse coordinate transformation.
This maps the discretization 
\begin{equation}
z^N_n=f(2y^N_n-1), \quad {\rm i.e.}, \quad  z^N_n=(2y^N_n-1)/\sqrt{1-q}
\end{equation} 
and the discretization-width becomes 
$dz\equiv 2/(\sqrt{1-q}(N+1))$, which takes into account the 
the factor $2f'(2y^N_n-1)=2/\sqrt{1-q}$ from the change of variables.
Analogous to $\rho$ the discrete formulation of the $q$-Gaussian $G_q$ 
reads $F^N_n=(dz)^{-1}\binom{N}{n}r^N_n$ and
\begin{equation}
 \lim_{N\to\infty}\sum_n|F^N_n-G_q(z^N_n)|\to 0 \quad.
 \end{equation}
This is exactly the model which was heuristically found in \cite{RodriguezSchwammleTsallis2008}. 
There the model was presented  as
\begin{equation}
r^N_0(q)=\frac{(N+\nu-1)(N+\nu-2)\dots \nu}{(N+2\nu-1)(N+2\nu-2)\dots (2\nu)} \quad, 
\end{equation}

for integer values of $\nu$, which is obviously  a particular case of Eq. (\ref{result1}). 

Since the $q$-Gaussian has no compact support for $q>1$ the situation becomes 
more involved, since now one has to map the real axis to the unit interval $[0,1]$. 
Such a $q$-Gaussian $G_q(z)$ with $z\in[-\infty,\infty]$ 
may be thought of as distributions of the distance $z$
a peculiar random walk has covered in the long time limit.
There is no map $f$ now that is a  natural choice, a priori.
However, in order to explicitly compute the probabilities $r^N_n$ one may choose a map
such that $r^N_n$ is given in terms of Beta functions, as before.
This leads to the situation that pairs of $q$-Gaussians, 
one with $q<1$ and another with $q>1$, can be generated   
by the same binary process and differ only in terms of the representation $f$.

Let us find a map $f$ such that $f(2y-1)$ again maps $y\in[0,1]$ to $z\in[-\infty,\infty]$.
Using the normalization condition of the $q$-Gaussian  this variable transformation implies  
$1=\int_{-\infty}^\infty dz\,G_q(z)=\int_{-1}^1 dx\,f'(x)G_q(f(x))$
and identify
\begin{equation}
  g(y)=2f'(2y-1)G_q(f(2y-1))\quad.	
\label{gxq2}
\end{equation}
A particular $f$ to compute $r^N_n$ in closed form is given by 
\begin{equation}
  f(y)=\frac{y}{\sqrt{1-y^2}}\frac{1}{\sqrt{q-1}} \quad {\rm and} \quad   f'(y)=\frac{\left(1-y^2\right)^{-3/2}}{\sqrt{q-1}} .
\label{model}
\end{equation}	
Noting, that this model $f$ for the $q$-Gaussian implies
$[1-(1-q)f(y)^2]^{1/(1-q)}=(1-y^2)^{1/(q-1)}$. Inserting this into Eq. (\ref{gxq2}) 
we finally get the stochastic model, 
\begin{equation}
r^N_n(q)=\frac{B\left(\mu+n,\mu+N-n\right)}{B\left(\mu,\mu \right)}\quad,
 \label{result2}
\end{equation}
where we have used the notation $\mu = \frac{1}{q-1}-\frac{1}{2}$. 
All that is left is to place everything correctly.
We use the same discretization $y^N_n=(1/2+n)/(N+1)$ for the interval $[0,1]$ as for $q<1$. 
Again, $dy^N_n=1/(N+1)$ is the width of the discretization on $[0,1]$ and with
$\rho^N_n=(dy^N_n)^{-1}\binom{N}{n}r^N_n$, we get that $\lim_{N\to\infty}\sum_n|\rho^N_n - g(y^N_n)|=0$, as a power of $N$
as in the $0<q<1$ case. 
To retrieve the  $q$-Gaussian $G_q(z)$, one has to perform the inverse coordinate transformation 
$y\in[0,1]\to z\in[-\infty,\infty]$. This is a little more complicated since the discretization width now 
depends on $N$ and $n$. 
In particular, mapping back to $[-\infty,\infty]$ gives the discretization 
$z^N_n=f(2y^N_n-1)$, i.e. $z^N_n=(y^N_n-1/2)/\sqrt{(q-1)y^N_n(1-y^N_n)}$, 
and the discretization width
becomes $dz^N_n\equiv (y^N_n(1-y^N_n))^{-3/2}/(4(N+1)\sqrt{q-1})$.
The discretized version of the $q$-Gaussian for $q>1$ now reads,
$F^N_n=(dz^N_n)^{-1}\binom{N}{n}r^N_n$. 
In the limit we get $\lim_{N\to\infty}\sum_n|F^N_n-G_q(z^N_n)|\to 0$, as a power of $N$.
Comparing Eq. (\ref{result1}) and Eq. (\ref{result2}) it is obvious that
for the two models of the $q$-Gaussian distribution ($q<1$ and $\bar q>1$) 
the generating binary processes $r^N_n$ are identical
whenever $\nu=\mu$.  $\nu$ and $\mu$ are functions of $q$ and $\bar q$, and 
 $\nu(q)=\mu(\bar q)$ establishes a relation between two $q$-Gaussian distributions 
generated by an identical exchangeable binary stochastic process, i.e. $\bar q=\frac{7-5q}{5-3q}$; $q$ increasing from $-\infty$ to 1 yields $\bar q$ decreasing from $5/3$ to 1. 
Therefore, the
models of $q$-Gaussian distributions with $q<1$ are conjugate with the model of 
$\bar q$-Gaussian with $\bar q\in[1,5/3]$ in the sense of being driven by an identical binary 
stochastic process.
The class of $\bar q$-Gaussian distributions with $\bar q\in(5/3,3]$, which
is exactly the class of normalizable $\bar q$-Gaussian distributions with diverging second moment,
are not  identified with a $q<1$. The corresponding binary processes are unique in this sense.
For instance, choosing $1<\bar q=2$ requires $1>q=3$, which is impossible.
For $\bar q=2$ it follows that $\mu=1/2$ and the associated binary process 
$r^N_n=B\left(1/2+n,1/2+N-n\right)/B\left(1/2,1/2 \right)$ has no representation for any $q<1$, and 
$r^N_0=\frac{B(1/2,1/2+N)}{B\left(1/2,1/2 \right)} 
=2^{-2N}\binom{2N}{N}$.
Of course one can choose many  families $(f_q,g_q)$ of models for $q$-Gaussian limit distributions
and, for each strictly monotonous function $\bar q:[-\infty,1]\mapsto[1,3]$, 
it is possible to construct {\em conjugate} families of models in the sense that
they are generated by the same family of binary processes.
In fact when the function $L_q$ is defined by $\int_0^{L_q(z)}dz' G_q(z')=\int_0^z dz''G_{\bar q(q)}(z'')$
and $(f_q,g_q)$ is a model for a $q$-Gaussian with $q<0$ then $(L_q(f_q), g_q)$ is a
conjugate model of the $\bar q$-Gaussian with $\bar q=\bar q(q)$. This allows to conjugate families of models
for different dualities recognized in $q$-statistics, e.g. 
$\bar q = (5-3q)/(3-q)$ for $q$-Fourier transforms \cite{nelsonumarov}.

We have explicitly derived  two possible stochastic models  for correlated 
and exchangeable  binary random variables, which lead to exact $q$-Gaussians as the 
limiting distributions,  while {\em strictly} satisfying Leibniz rule, 
\begin{equation}
r_n^N  = \left\{
\begin{array}{llll}
\dfrac{B\left(\frac{2-q}{1-q}+n,\frac{2-q}{1-q}+N-n\right)}{B\left(\frac{2-q}{1-q},\frac{2-q}{1-q} \right)}   
 &&&\textrm{if}\quad q<1 \, , \\  
    1/2^N &&&\textrm{if} \quad q=1 \,,      \\   
   \dfrac{B\left(\frac{3-q}{2q-2}+n,\frac{3-q}{2q-2}+N-n\right)}{B\left(\frac{3-q}{2q-2},\frac{3-q}{2q-2} \right)}  
&&& 
\textrm{if} \quad 1<q <3 \,,  
\end{array}\right.
\label{eq14}
\end{equation}
$q=1$ can be obtained by both $q\to1^\pm$. 
Eq. (\ref{eq14}) 
follows for the affine representation for
$q<1$ and for the representation 
$f$ given in Eq. (\ref{model}) for $q>1$. 
For different families 
of representations $f_q$ of the q-Gaussians the equation has to be adapted
accordingly.

\begin{figure}[t]
\includegraphics[width=0.50\textwidth]{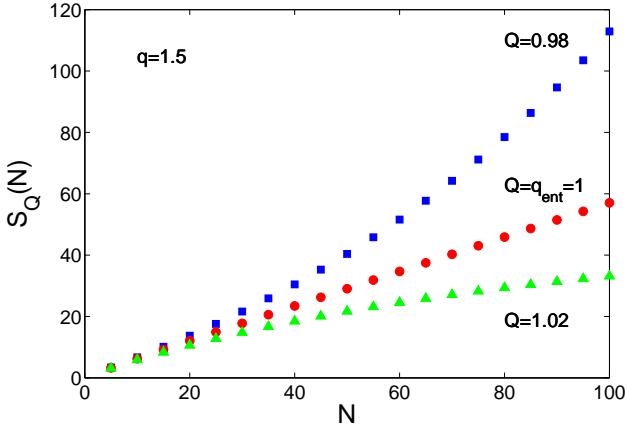}
\caption{Nonadditive entropy $S_Q$ vs. $N$  for a $q$-Gaussian with $q=1.5$, for $Q=0.92,1$ and $1.02$. Clearly, only  
$Q =1$ is extensive (i.e., $S_1(N) \propto N$, for $N\gg1$). 
}
\end{figure}

Lately the question whether $S_Q$ can be the extensive entropy for specific systems for $Q<1$ has been discussed, e.g. in \cite{TsallisGellMannSato2005,CarusoTsallis2008}. 
With the Laplace-de Finetti representation we consider  
the entropy  $S_Q$ for large $N$ and prove that the value of $Q$ for which $S_Q$ is extensive with respect
to the distribution of binary states is $Q=1$, i.e. $q_{ent}=1$, by straight foreward calculation: 
\begin{eqnarray}
&&S_{Q}[g] = \frac{1-\sum_{n=0}^{N}  \binom{N}{n} [r^N_n]^{Q}   }{Q-1} \;\;\;\;\;\;\;\; \nonumber \\
&=&\frac{\sum_{n=0}^{N}\frac{\left( (N+1)\binom{N}{n} \right)^{1-Q}}{N+1}  \left[\frac{\int_0^1 dx\, x^n(1-x)^{N-n}g(x)}{\int_0^1 dx\, x^n(1-x)^{N-n}}\right]^{Q} -1  }{1-Q} \;\;\;\;\;\;\;\; \nonumber \\
     &=& \frac{    \int_0^1  dy \, \left[ \int_0^1 dx \,  [x^y (1-x)^{1-y}]^N  \right]^{Q-1}       [g(y)]^{Q} -1  }{1-Q}\,. \;\;\hspace{0.5cm}
\end{eqnarray}
The first line follows from using the exchangeability-property together with the definition of the 
$q$-entropy. The second line can be obtained by inserting Eq. (\ref{definetti}) for $r^N_n$ and using the
properties of the $\beta$-function. The third line follows for large $N$, where 
$\sum_{n=0}^N 1/(N+1)\to \int_0^1 dy$, and using Eq. (\ref{eqb}).
To find the index $Q$ where $S_Q$ gets extensive requires $\frac{d S_{Q}[g]}{dN}$ to be a positive constant for $N\gg1$ for
some specific value of $Q$ which is called $q_{\rm ent}$. 
We find, by using Stirling's approximation, 
\begin{eqnarray}
  \frac{d \, S_{Q} [g](N)}{dN} 
 &\sim& 2^{N(1-Q)} \, g \left( 1/2 \right) \ln 2 \quad .
\end{eqnarray}
Hence, for any $g(1/2)>0$, 
$S_Q(N)$ growing linearly with large $N$ requires that $Q=1$,
since otherwise $S_Q(N)$ grows exponentially with $N$, 
i.e., $q_{\rm ent}=1$ for all such models.  
It should be noted that 
particular choices of representations $f$ and the observable process (distribution functions) only
result in particular choices of $g$. Since the result is valid for all $g$ the result is basically
independent of the particular choice of representing an exchangeable process.

We have shown that the Laplace-de Finetti representation 
is a suitable framework which allows to generate scale invariant probabilistic models by 
fixing their limit distribution and a map which describes how the domain of the observable variable
of model is embedded on the real axis on each scale $N$. We have demonstrated this with
the $q$-Gaussian distributions as an example. We have constructed scale invariant models
for $q$-Gaussians for all $q<3$ and have shown how notions of conjugate families of
processes can be constructed. We have shown that the Boltzmann Gibbs entropy is the extensive entropy for
all stochastic processes generated by scale-invariant binary processes.

We acknowledge useful remarks by A. Rodriguez, V. Schwammle, E.P. Borges and E.M.F. Curado. 
R.H. and S.T. acknowledge  great hospitality at CBPF and  partial support by Faperj and CNPq (Brazilian agencies) 
and FWF Project P19132.


\begin{thebibliography}{99}

\bibitem{AnteneodoTsallis2003} Anteneodo C. and Tsallis C., 
J. Math. Phys., {\bf 44}  (2003) 5194 .

\bibitem{FuentesCaceres2008} Fuentes M.A. and  Caceres M.O.,  
Phys. Lett. A, {\bf 372} (2008) 1236-1239.

\bibitem{Borland1998} Borland L., 
Phys. Lett. A, {\bf 245} (1998) 67.

\bibitem{PlastinoPlastino1995} Plastino A.R. and Plastino A., 
Physica A,  {\bf 222} (1995) 347.

\bibitem{TsallisBukman1996}Tsallis C. and  Bukman D.J., 
Phys. Rev. E, {\bf 54} (1996) R2197.

\bibitem{UmarovTsallisSteinberg2008}Umarov S., Tsallis C. and Steinberg S., 
Milan J. Math., {\bf 76} (2008) [DOI 10.1007/s00032-008-0087-y]; for a more pedagogical version, see  Queiros S.M.D. and Tsallis C., 
AIP {\bf 965}, 21 (New York, 2007).


\bibitem{Tsallis1988} Tsallis C., 
J. Stat. Phys., \textbf{52} (1988) 479; 
{\it Introduction to Nonextensive Statistical Mechanics - Approaching a Complex World} (Springer, New York, 2009), in press.


\bibitem{Penrose1970}Penrose O., {\it Foundations of Statistical Mechanics: A Deductive Treatment} (Pergamon, Oxford, 1970), p. 167.

\bibitem{TsallisGellMannSato2005}Tsallis C., Gell-Mann M. and Sato Y., Proc. Nat. Acad. Sci., \textbf{102} (2005) 15377.

\bibitem{CarusoTsallis2008}Caruso F. and Tsallis C., 
Phys. Rev. E, {\bf 78}  (2008) 021101.



\bibitem{PluchinoRapisardaTsallis2007}
Pluchino A., Rapisarda A. and Tsallis C., \textit{Eur. Phys.
Lett.}, \textbf{80} (2007) 26002; Physica A, {\bf 387} (2008) 3121.



\bibitem{AnteneodoTsallis1998} Anteneodo C. and Tsallis  C., Phys. Rev. Lett., \textbf{80} (1998) 5313.

\bibitem{yamaguchi}
Yamaguchi Y., Bouchet G. and Dauxois T.,
JSTAT, (2007) P01020.  

\bibitem{bouchet}
Bouchet G.  and T. Dauxois T., 
Phys. Rev. E, {\bf 72} (2005) 045103 (R). 

\bibitem{UpadhyayaRieuGlazierSawada2001}Upadhyaya A.,  Rieu J.-P.,  Glazier J.A. and  Sawada Y., 
Physica A, {\bf 293} (2001) 549.

\bibitem{cellmove}
Thurner S., Wick N.,  Hanel R.,  Sedivy R. and  Huber L.A., 
Physica A, {\bf 320C}  (2003) 475-484.

\bibitem{DanielsBeckBodenschatz2004} Daniels K.E.,  Beck C. and  Bodenschatz E., 
Physica D, {\bf 193}  (2004) 208.

\bibitem{BurlagaVinas2005} Burlaga L.F. and Vinas A.F., 
Physica A, {\bf 356}  (2005) 375.

\bibitem{BurlagaVinasAcuna2006} Burlaga L.F.,  Vinas A.F.,  Ness N.F. and  Acuna M.H.,  
Astrophys. J., {\bf 644}  (2006) L83-L86.

\bibitem{DouglasBergaminiRenzoni2006} Douglas P.,  Bergamini S. and Renzoni  F., 
Phys. Rev. Lett., {\bf 96}  (2006) 110601.

\bibitem{LiuGoree2008} Liu B. and  Goree J., 
Phys. Rev. Lett., {\bf 100}  (2008) 055003.

\bibitem{TirnakliBeckTsallis2007}Tirnakli U.,  Beck C. and  Tsallis C., 
Phys. Rev. E, {\bf 75} (2007) 040106(R);  0802.1138 [cond-mat.stat-mech]. 


\bibitem{grass}
Grassberger P., 
arXiv:0809.1406  [cond-mat.stat-mech]. 

\bibitem{RodriguezSchwammleTsallis2008} Rodriguez A.,  Schwammle V. and  Tsallis C., 
JSTAT (2008) P09006.

\bibitem{definetti}
de Finetti B.,
Atti della R. Academia Nazionale dei Lincei 6, 
Memorie, Classe di Scienze Fisiche, Mathematice e Naturale, 
{\bf 4}  (1931) 251Ð299;
Heath D.,  Sudderth W., 
Am. Stat.,  {\bf 30}  (1976) 188-189. 

\bibitem{jaynes}
 Jaynes E.T., 
in {\it Bayesian Inference and Decision Techniques with Applications},
 Goel P.K. and  Zellner A., eds. (North-Holland, Amsterdam, 1986), p. 31.

\bibitem{hewittsavage}
 Hewitt E.,  Savage L.J.,  
Trans. Am. Math. Soc., {\bf 80} (1955)  470-501. 

\bibitem{nelsonumarov}
 Nelson K.P., Umarov S.,
arXiv:0811.3777v1.

\end{thebibliography}
\end{document}